\documentclass{article}

\usepackage{PRIMEarxiv}

\usepackage[utf8]{inputenc} 
\usepackage[T1]{fontenc}    
\usepackage{hyperref}       
\usepackage{url}            
\usepackage{booktabs}       
\usepackage{amsfonts}       
\usepackage{nicefrac}       
\usepackage{microtype}      
\usepackage{lipsum}
\usepackage{amsmath}
\usepackage{fancyhdr}       
\usepackage{graphicx}       
\graphicspath{{media/}}     

\title{Scalable Quantum Ground State Preparation of the Heisenberg Model: A Variational Quantum Eigensolver Approach
}

\author{
  Jinao Wang\\
  Tsinghua International School\\
  Beijing, China\\
  \texttt{michaelwja.0114@gmail.com} \\
   \And
  Rimika Jaiswal \\
  Department of Physics \\
  University of California, Santa Barbara \\
  Santa Barbara\\
}

\begin{document}
\maketitle
\begin{abstract}
Quantum systems have historically been formidable to simulate using classical computational methods, particularly as the system size grows. In recent years, advancements in quantum computing technology have offered new opportunities for tackling complex quantum systems, potentially enabling the study and preparation of quantum states directly on quantum processors themselves. The Variational Quantum Eigensolver (VQE) algorithm is a system composed of a quantum circuit as well as a classical optimizer that can be used to efficiently prepare interesting many-body states on the current noisy intermediate-scale quantum (NISQ) devices. We assess the efficacy and scalability of VQE by preparing the ground states of the 1D generalized Heisenberg model, a pivotal model in understanding magnetic materials. We present an ansatz capable of preparing the ground states for all possible values of the coupling, including the critical states for the anisotropic XXZ model. This paper also aims to provide insights into the precision and time consumption involved in classical and optimized sampling approaches in the calculation of expectation values. In preparing the ground state for the Heisenberg models, this paper paves the way for more efficient quantum algorithms and contributes to the broader field of condensed matter physics.

\end{abstract}


\section{Introduction}
\par Recent advancements in quantum computing have enabled researchers to apply quantum circuits in analyzing the quantum states of complex many-body systems, which have been previously formidable to classical computational approaches. When it comes to quantum state preparations, classical computation methods face significant scalability issues as the computational cost grows exponentially in the Hilbert Space \cite{carleo_nomura_imada_2018}. In addition, as stated by John Preskill, the current level of quantum technology is best characterized by Noisy Intermediate Scale Quantum (NISQ) computing that has high error rates and limited system size \cite{preskill_2018}
\par In order to address the limits posed by current quantum hardware, a class of algorithms called Variational Quantum Algorithms was proposed. VQE, a subclass of VQAs, will be closely examined in this paper. First proposed by Alberto Peruzzo et.al \cite{peruzzo_mcclean_shadbolt_yung_zhou_love_aspuru-guzik_o’brien_2014},  VQE is a hybrid quantum-classical algorithm that has emerged as a promising approach for solving problems in quantum chemistry, material science, and condensed matter physics using near-term quantum computers. VQE is an application of the variational principle, where a quantum computer is trained to prepare the ground state of a given system, such as a molecule or a spin lattice \cite{Tilly_2022}.

\par The main goal of this paper is to use VQE to prepare the ground state of the generalized Heisenberg model variationally. The Heisenberg model is a fundamental theoretical framework in the study of magnetic systems and plays a crucial role in understanding the behavior of quantum spin systems. Developed by Werner Heisenberg, this model describes the interactions between magnetic moments or spins in a lattice, taking into account their quantum mechanical nature \cite{cerezo_arrasmith_babbush_benjamin_endo_fujii_mcclean_mitarai_yuan_cincio_et_al_2021}. By determining the ground state of the Heisenberg model, which corresponds to the lowest energy configuration of the system, physicists can better understand its low-temperature properties and phase transitions, as well as providing insights into the nature of quantum correlations and entanglement in the system \cite{mediumLookVQE}. 
\par Current attempts to use the Variational Quantum Eigensolver (VQE) to calculate the Heisenberg model's ground state have demonstrated promise and versatility, yet encounter significant challenges. Early research by Cervera-Lierta et al. and Kokail et al. validated VQE's potential, delivering accurate results for small systems but highlighting scalability issues due to limited qubits and quantum device noise \cite{ciavarella_caspar_illa_savage_2023,kokail_maier_van_bijnen_brydges_joshi_jurcevic_muschik_silvi_blatt_roos_et_al_2019}.

\par Moreover, in the study of the anti-ferromagnetic Heisenberg model on the kagome lattice, VQE results aligned with exact diagonalization for an 18-site system, except near phase transition points where it erroneously converged to excited states \cite{kattemolle_van_wezel_2022}. Moll et al. also applied VQE within quantum chemistry, again underscoring the difficulty of larger system optimization due to its complex landscape \cite{moll_barkoutsos_bishop_chow_cross_egger_filipp_fuhrer_gambetta_ganzhorn_et_al_2018}.
\par In tackling the scalability issues in ground state preparations using VQE, this paper focuses not only on obtaining the target ground state but also analyzing the computation efficiency of VQE and studying the effect of sampling optimization on the VQE runtime.  In addition to the widely explored 1D isotropic Heisenberg model for small lattices, this paper also studies larger systems with more than 20 qubits. The less explored XXZ anisotropic Heisenberg model is also examined in this paper, demonstrating VQE’s capability in a variety of models.   

\section{Methods}
\label{sec:headings}
\subsection{The Variatinal Quantum Eigensolver}
Suppose we want to prepare a target ground state $ \vert \psi_{target} \rangle $.  The VQE takes in a trivial input state $ \vert \psi_{0} \rangle $ and finds the appropriate parametrized unitary circuit that can act on the input state to generate the target state. The optimization process is implemented on a hybrid system of classical optimizer and quantum circuit, where a set of parameters are given to a quantum circuit, and the expectation values are calculated, then sent back to the classical optimizer where the optimizer updates a new set of parameters to the quantum circuit.

\begin{figure}[h]
    \centering
    \includegraphics[width=0.5\textwidth]{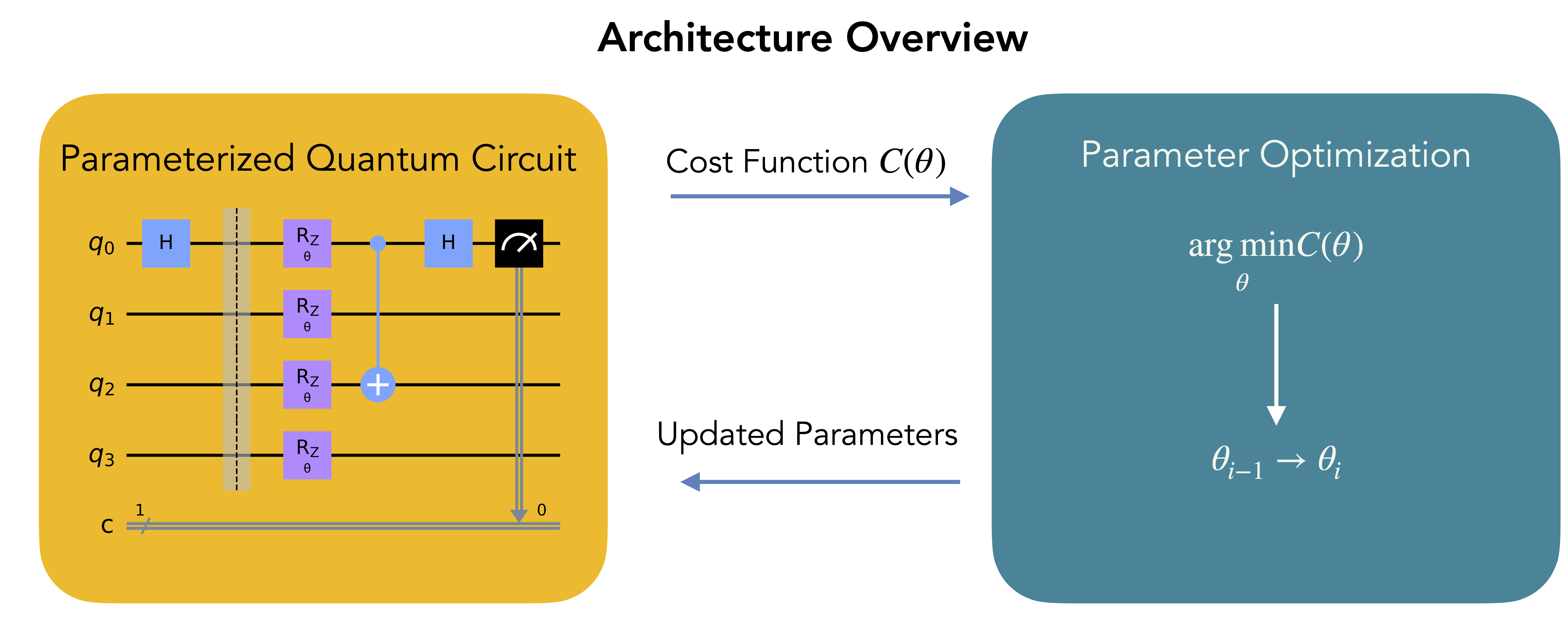}
    \caption{Schematic overview of the structure of the Variational Quantum Eigensolver (VQE).}
    \label{fig:arch_overview}
\end{figure}

\subsection{1D Isotropic Heisenberg Model}

\label{sec:2b}

The Hamiltonian for the 1D Heisenberg model is defined as
\begin{equation}
H_{T}=J\sum_{i=1}^{L-1} \boldsymbol{S}_{i} \cdot \boldsymbol{S}_{i+1}
\end{equation}
where S represents the spin operator of a particle, and J is the coupling constant that will be defined as 1 for the isotropic model.
As for the ansatz design, this paper follows the quantum computing simulation protocol devised by Wen Wei Ho and Timothy H. Hsieh [6]. The final target state is written as a linear combination of  and , which will be defined as $H_1=\sum_{i=1}^{L / 2-1} \boldsymbol{S}_{2 i} \cdot \boldsymbol{S}_{2 i+1}$ , and $H_2=\sum_{i=1}^{L / 2} \boldsymbol{S}_{2 i-1} \cdot \boldsymbol{S}_{2 i}$ . Then the structure of the ansatz can be represented as alternating layers of $H_1$ and $H_2$ with $p \in \mathbb{Z}$ parameters for $  \beta $ and  $\gamma $ respectively: 

\begin{equation}
|\psi(\boldsymbol{\gamma}, \boldsymbol{\beta})\rangle_p=e^{-i \beta_p H_1} e^{-i \gamma_p H_2} \cdots e^{-i \beta_1 H_1} e^{-i \gamma_1 H_2}\left|\psi_1\right\rangle
\end{equation}

\par
\begin{figure}[t]
    \centering
    \includegraphics[width=0.75\textwidth]{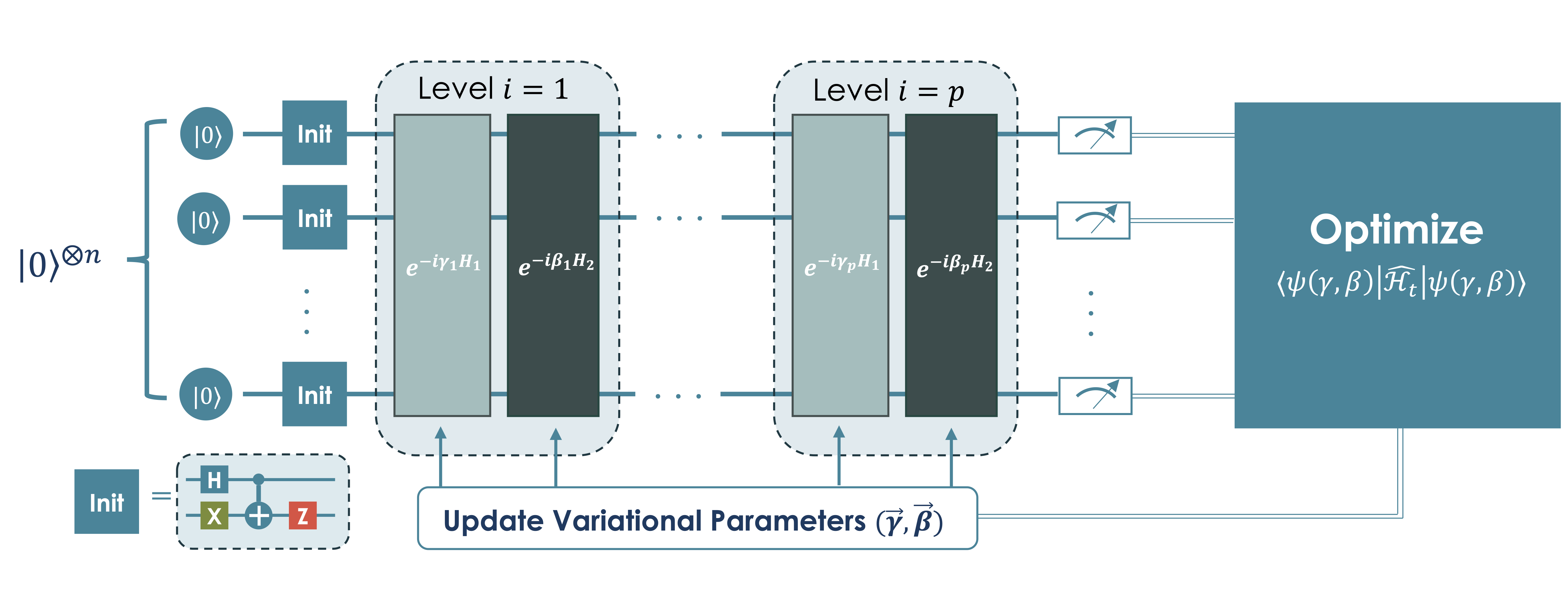}
    \caption{Schematic overview of the structure of the ansatz.}
    \label{fig:schematic_design}
\end{figure}

 Figure\ref{fig:schematic_design} shows the detailed layout of the ansatz design for the 1D Heisenberg model. The circuit is first initialized into entangled bell state pairs $\bigotimes_i \frac{1}{\sqrt{2}}(|\uparrow \downarrow\rangle-|\downarrow \uparrow\rangle)_{2 i-1,2 i}$ . Then, parameterized $H_1$ and $H_2$ layers alternate to form $p$ levels, with each level containing two layers of operation gates. This design of alternating H1 and H2 layers is adopted from the Quantum Approximate Optimization Algorithm (QAOA), which is originally developed to solve combinatorial optimization problems \cite{RN31}. This ansatz design is effective for certain problems that have a feasible subspace that is smaller than the full Hilbert space \cite{cerezo_arrasmith_babbush_benjamin_endo_fujii_mcclean_mitarai_yuan_cincio_et_al_2021}.  
\par The expectation value with respect to the Hamiltonian (also the cost function)is efficiently defined as:
\begin{equation}
F_p(\gamma, \beta)={ }_p\left\langle\psi(\gamma, \beta)\left|H_T\right| \psi(\gamma, \beta)\right\rangle_p
\end{equation}

\subsection{1D Anisotropic XXZ Heisenberg Model}
The XXZ model is a generalization of the 1D Heisenberg model given in the form below
\begin{equation}
H=\sum_{i=1}^N\left(J_x S_i^x S_{i+1}^x+J_y S_i^y S_{i+1}^y+J_z S_i^z S_{i+1}^z\right)
\end{equation}
that satisfies the condition $J_x=J_u \neq J_z$. In this paper, $ J_x=1 $ sets the energy scale and $ J_z = \Delta $  is the anisotropy parameter. Although the XXZ model is already exactly solvable with the Bethe ansatz \cite{miao_lamers_pasquier_2021, samaj_bajnok_2013}, its ground state preparation has not been explored using the VQE.
\par The ansatz of the XXZ model is similar to the one used in the calculation for the 1D Heisenberg model (Figure \ref{fig:schematic_design}). But here, the anisotropy parameter $\Delta $needs to be passed into both the Hamiltonian and the variational circuit. When $ \Delta=0 $, the model becomes the XY model, neglecting the spin interaction in the Z-axis [19]. When $ \Delta=1 $, the model recovers isotropy with an $SU(2)$ symmetry \cite{yang_wang_wu_2023}. For $ \Delta \in [-1, 1]$, the model is in the critical phase, which will be selected as the value range in the subsequent experiments \cite{amslaurea9341}.

\subsection{Sampling Optimization}  
\par In addition to the VQE, sampling is also applied in the calculation of the expectation value to further reduce the computational cost as $N$ increases \cite{libretexts25Expectation}. Sampling utilizes the probabilistic nature of quantum mechanics by randomly taking a subset of all the possible outcomes, instead of calculating all the possible states. In doing so, the task of calculating the expectation value is reduced from exponential complexity down to polynomial complexity, with an error of $\frac{1}{\sqrt{M}}$  given by the central limit theorem \cite{cushen_hudson}, where $M$  denotes the number of samples \cite{rubin_babbush_mcclean_2018,RN16}. 

\section{Results and Discussions}
The ground state of  the aforementioned models is prepared for different system sizes. To demonstrate the scalability of VQE with sampling optimization, ground states computations are also conducted on larger lattices and analyzed by recording the computation time. 

\subsection{1D Isotropic Heisenberg Model }

\begin{figure}[h]
    \centering
    \includegraphics[width=0.5\textwidth]{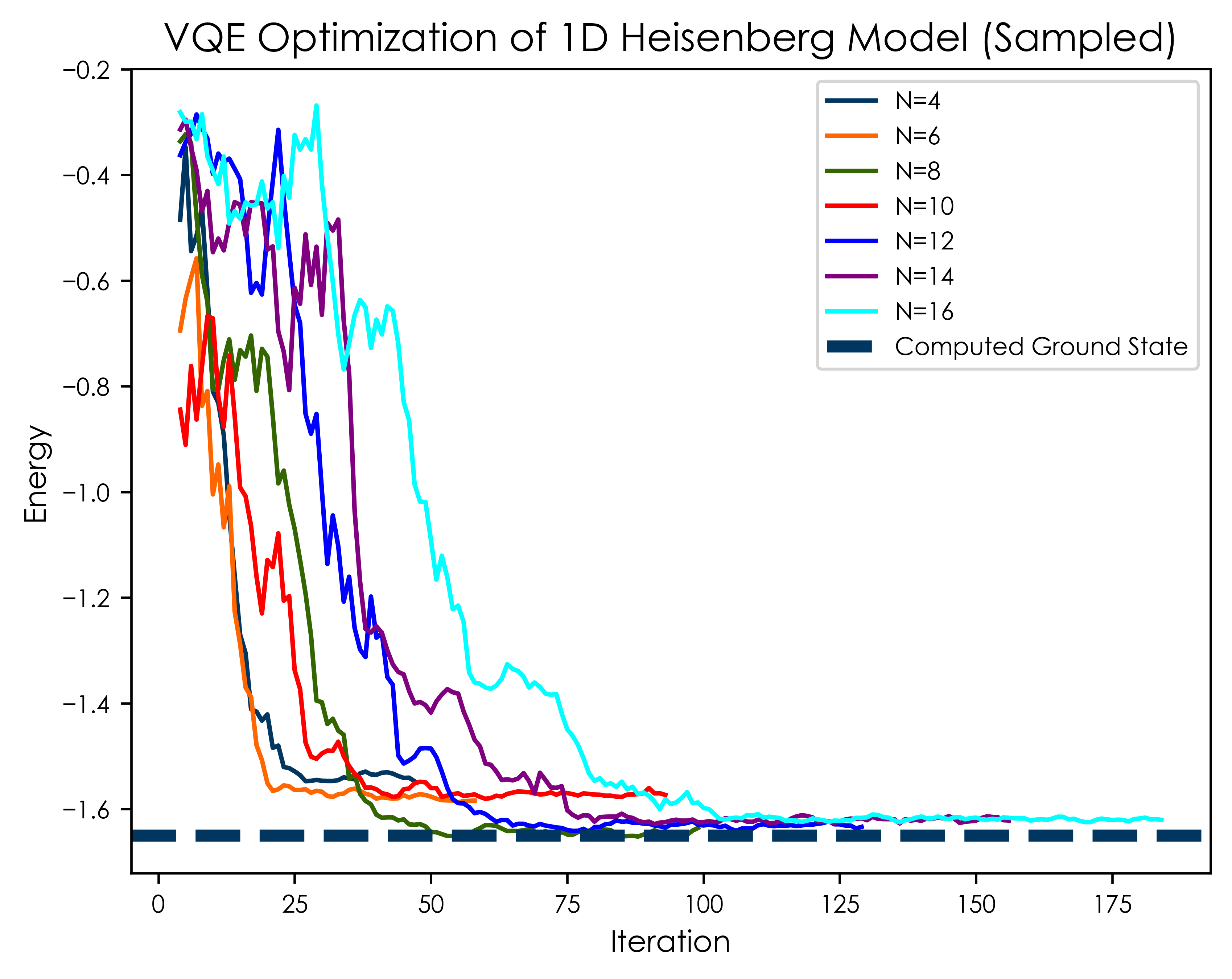}
    \caption{Optimization record of 1D Heisenberg Model for N = 4,6,8,10,12,14,16, with sampling optimization applied. }
    \label{fig:1d-isotropic-sample}
\end{figure}

In Figure \ref{fig:1d-isotropic-sample}, the optimization was performed on 1D isotropic Heisenberg model with varying system sizes. In most cases, the calculated ground state converges to the correct ground state, but in some cases, such as N = 12 and N = 6, the optimization energy does not converge exactly to the target energy. This is likely due to the fact that the use of sampling for the calculation of the ground state led to more randomized results, while in FIG 9 of Appendix 1A, the prepared ground state through exact calculation of the expectation value demonstrates higher accuracy. 

\subsection{1D Anisotropic Heisenberg Model}

\begin{figure}[h]
    \centering
    \includegraphics[width=0.5\textwidth]{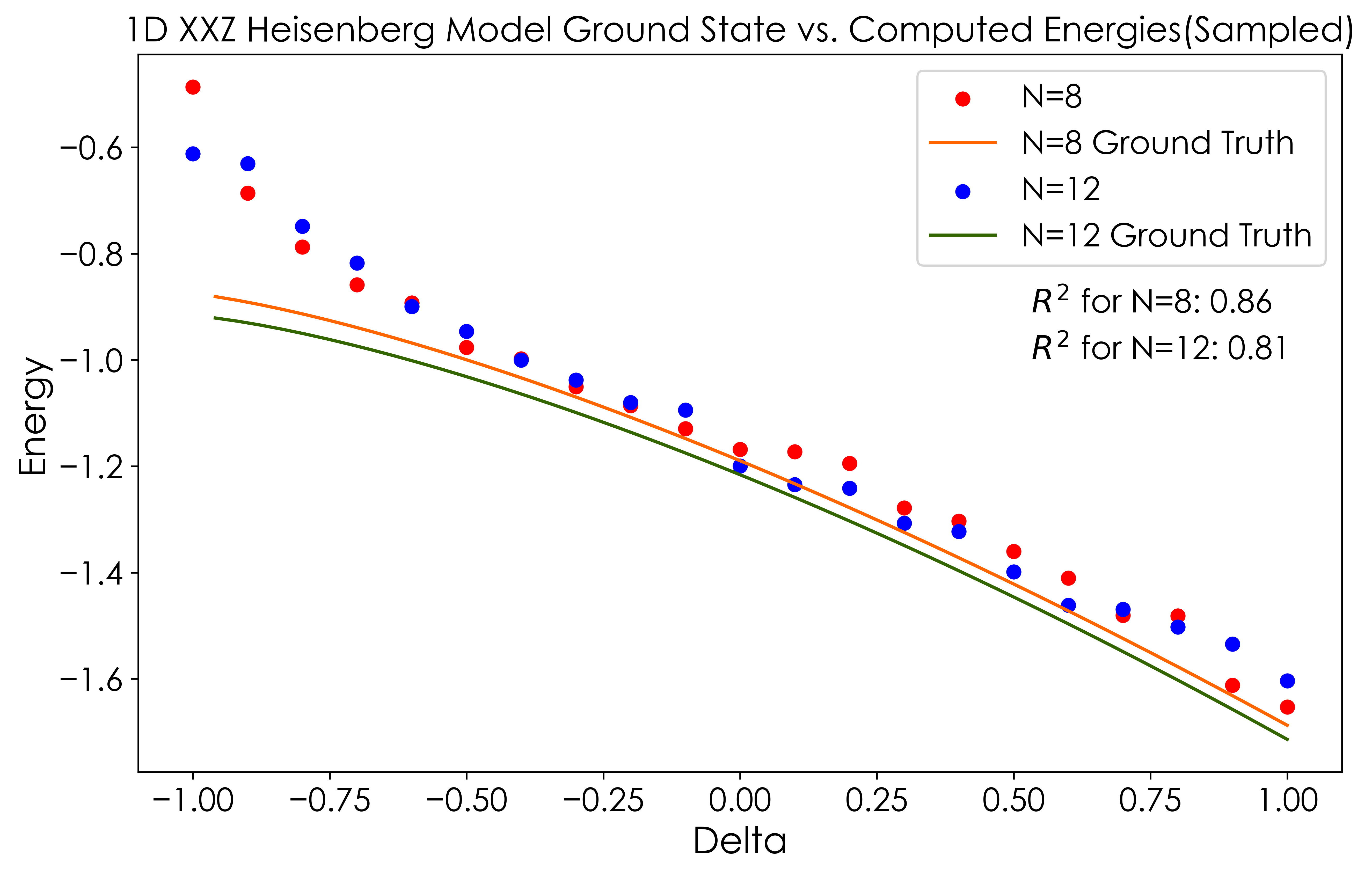}
    \caption{Ground states of the 1D Anisotropic XXZ Heisenberg model with $\Delta \in [-1,1]$ sampled at an interval length of 0.1, sampling optimization applied. }
    \label{fig:1d-anisotropic-sample}
\end{figure}

For the anisotropic XXZ model, the ground state is prepared for a set of delta values belonging to the critical phase of the XXZ model exhibiting the characteristics discussed in section \ref{sec:2b}. The sampling results are then compared with the calculated values obtained through exact diagonalization shown in figure \ref{fig:1d-anisotropic-sample}. The $ R^2 $ of both runs calculated with respect to the computed energy is above 0.8, indicating a strong correlation between the optimized value and the ground truth, thereby demonstrating the accuracy of VQE in preparing the ground state of the XXZ anisotropic models.

\subsection{Running on Larger Systems and Runtime analysis}

\begin{figure}[h]
    \centering
    \includegraphics[width=0.5\textwidth]{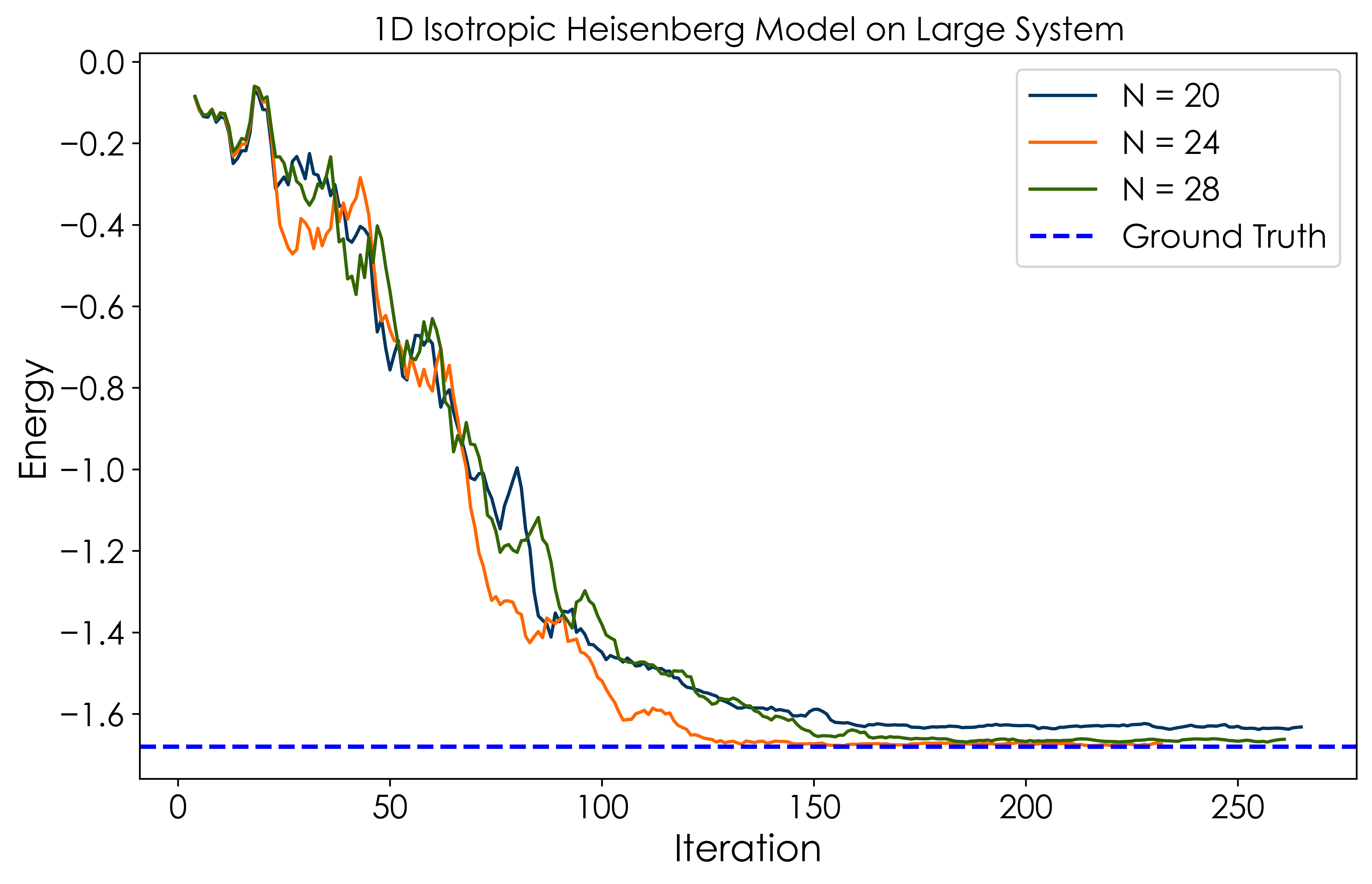}
    \caption{Optimization results of 1D Isotropic Heisenberg Model  for N = 20,24,28, with sampling optimization applied. }
    \label{fig:large_sys}
\end{figure}

\par For larger systems, the VQE still converges to the expected ground state for the three runs conducted, demonstrating its stability and robustness even when the system size increases.

\begin{figure}[h]
    \centering
    \includegraphics[width=0.5\textwidth]{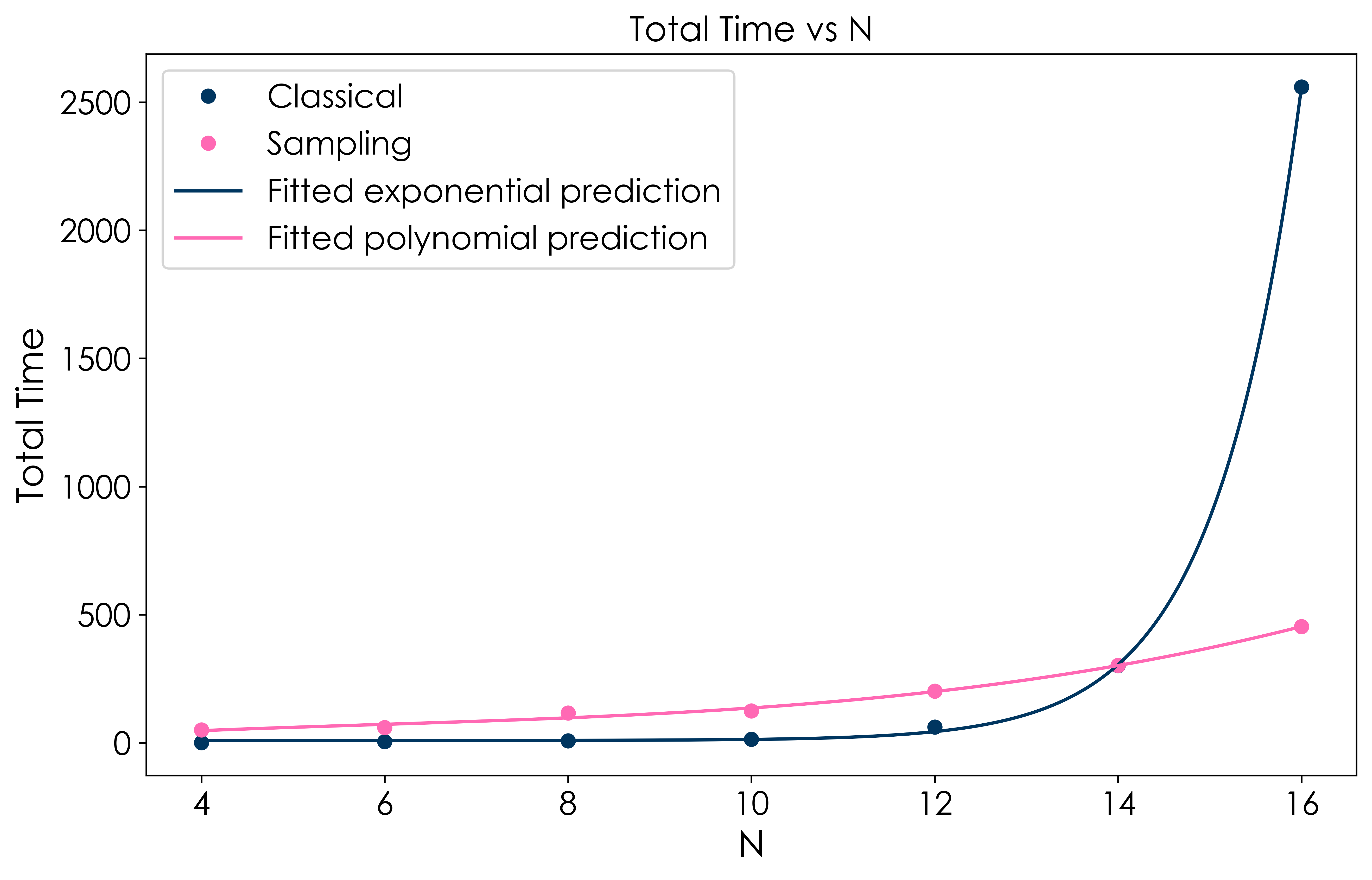}
    \caption{VQE optimization time comparison between exact computation and sampling optimization.}
    \label{fig:time_analysis}
\end{figure}

\par When comparing the computation time with and without sampling optimization for the calculation of the expectation value, classical computation is faster on smaller-scale systems. There is an overlap between the two approaches at N=14. But as N increases beyond 14, the runtime of the classically computed cost function increases significantly while the runtime for sampling increases much slowly (only polynomially)This shows the exponential growth in the computation complexity of expectation values classically and also proves the efficiency of sampling optimization that only has polynomial time complexity. 

\subsection{Result Verification}
To verify the result of the calculated ground state, the subsystem entropy, as well as the correlation function of one set of optimized parameters, is calculated. If the calculated results demonstrate the expected physical properties, then it can be confirmed that the results obtained through this research are valid. Due to the length constraint, only the result  for $\Delta=1$ is presented. For the full result, please refer to Appendix 2a and 2b.

\begin{figure}[h]
    \centering
    \includegraphics[width=0.75\textwidth]{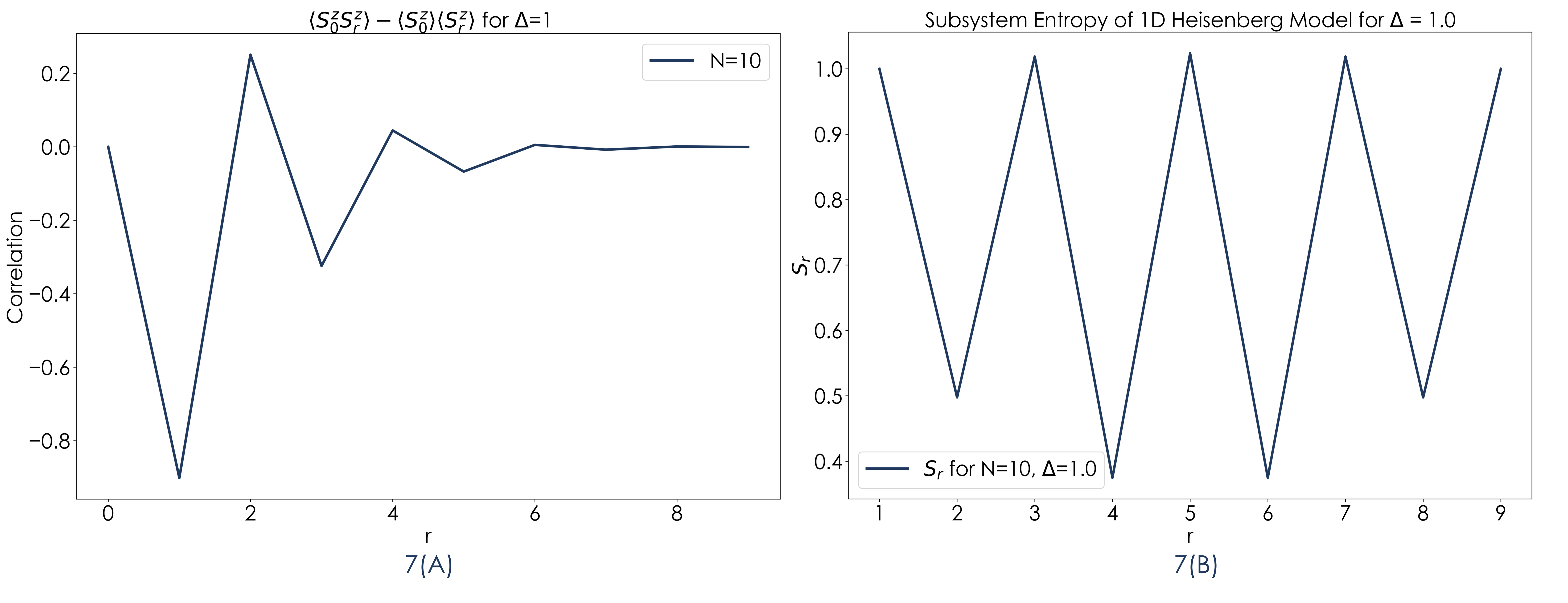}
    \caption{Result verification of the prepared ground state for $N=10$ and $ \Delta=1$. 7(a) the correlation  of the prepared state with respect to r(distance between qubits). 7(b) the subsystem entropy of the 1D Heisenberg.  }
    \label{fig:verification}
\end{figure}

Graph \ref{fig:verification}(b) shows an oscillating pattern, confirming our system's initialization into a dimer state with entangled Bell pairs. High entropy occurs when subsystem cuts intersect entangled qubits, while low entropy arises from unentangled pair intersections, confirming the dimerized structure. Graph \ref{fig:verification}(a) exhibits an oscillatory correlation function that starts strong at short distances, indicative of singlets, then rapidly drops to zero, typical of a strongly dimerized state for the Heisenberg model. Thus, by analyzing the graph’s characteristics, it can be concluded that the prepared state is valid. 

\section{Conclusion}
\par In this paper, the optimized parameters are obtained for the variational quantum circuit at different system sizes and both the isotropic and anisotropic models. This is valuable for future researchers who wish to obtain the ground state and its properties directly through variational quantum circuits. This can be directly useful in the simulations of magnetic systems. 

\par In addition, this paper also demonstrates the capabilities of the VQE for the XXZ model, which has not been implemented using VQE in the past. Based on the obtained optimization results, it can be concluded that VQE can reliably be used to prepare the ground state of the XXZ anisotropic Heisenberg model. 
\par Sampling optimization is verified to improve the time runtime from exponential complexity down to polynomial complexity while retaining most of the accuracy.
\par Lastly, VQE also demonstrates robustness as it can still reliably converge to the desired ground state even on large systems that are out of reach for classical computations. 

\section*{Acknowledgments}
\par I would like to express my heartfelt gratitude to my research partner, Lana Wong, for her support and encouragement throughout this research. A big thank you to my TA Mr. Vamshi Chowdary Madala for giving me exhaustive and detailed advice on my research project, and answering my questions and concerns with patience.  

\par Additionally, my sincere appreciation goes to the Program Director of the Research Mentorship Program (RMP) at UCSB, Dr. Lina Kim, for her wonderful lectures and support in crafting and presenting all crucial components of my research project.

\par Finally, I would like to thank all the UCSB faculty and those who helped organize RMP 2023. This paper would not have been possible without your collective support and mentorship.

\bibliographystyle{ieeetr}  
\bibliography{references}  

\newpage

\section*{Appendix}

\subsection{Exact Computation Optimization Records}
  The ground state optimization for the isotropic Heisenberg model is obtained via the exact computation of the expectation values. When compared with the results obtained from sampling, the results are more accurate but also need more iterations to converge.

\begin{figure}[h]
    \centering
    \includegraphics[width=0.75\textwidth]{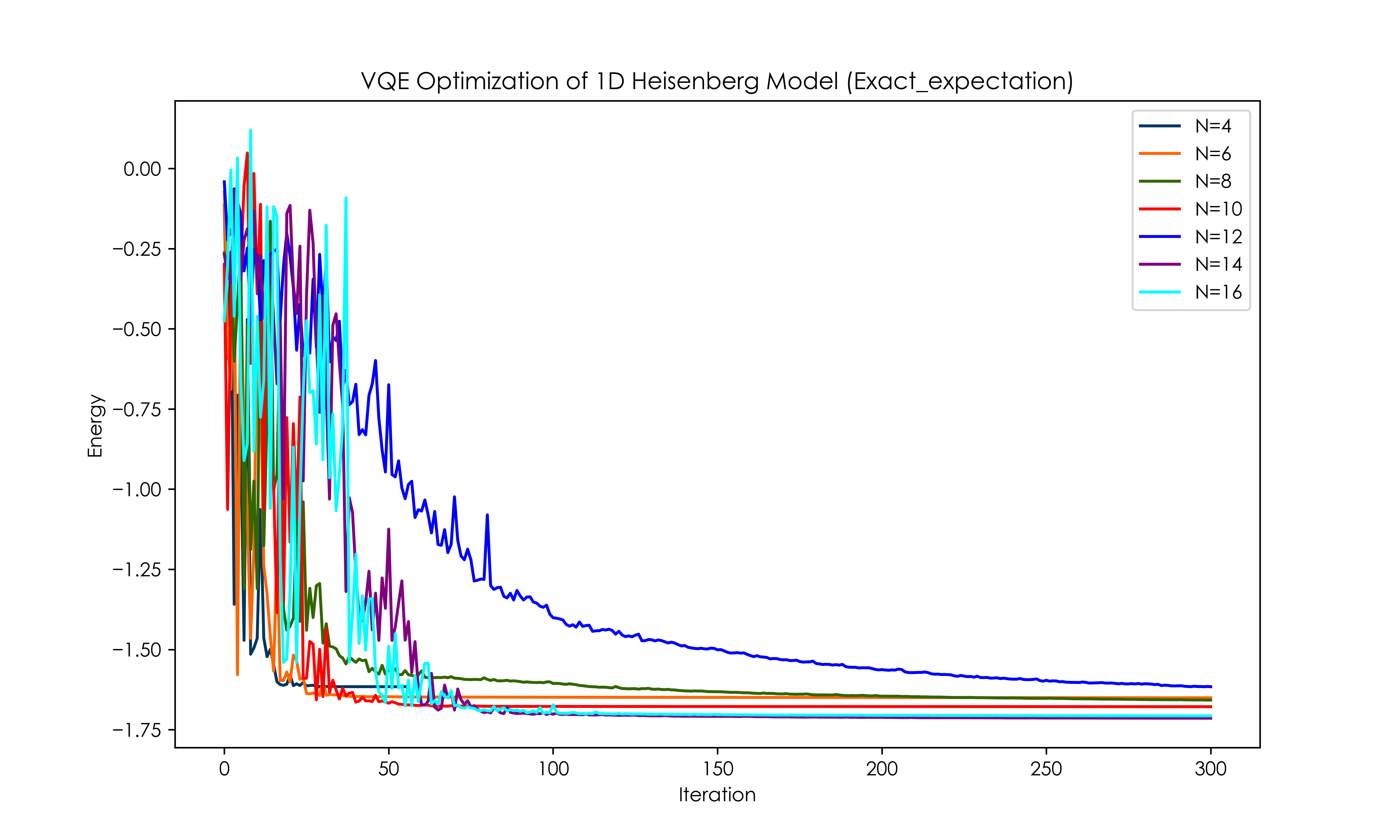}
    \caption{Optimization record of 1D Heisenberg Model for N = 4,6,8,10,12,14,16, with classically computed expectation value}
    \label{fig:apd1}
\end{figure}

\begin{figure}[h]
    \centering
    \includegraphics[width=0.75\textwidth]{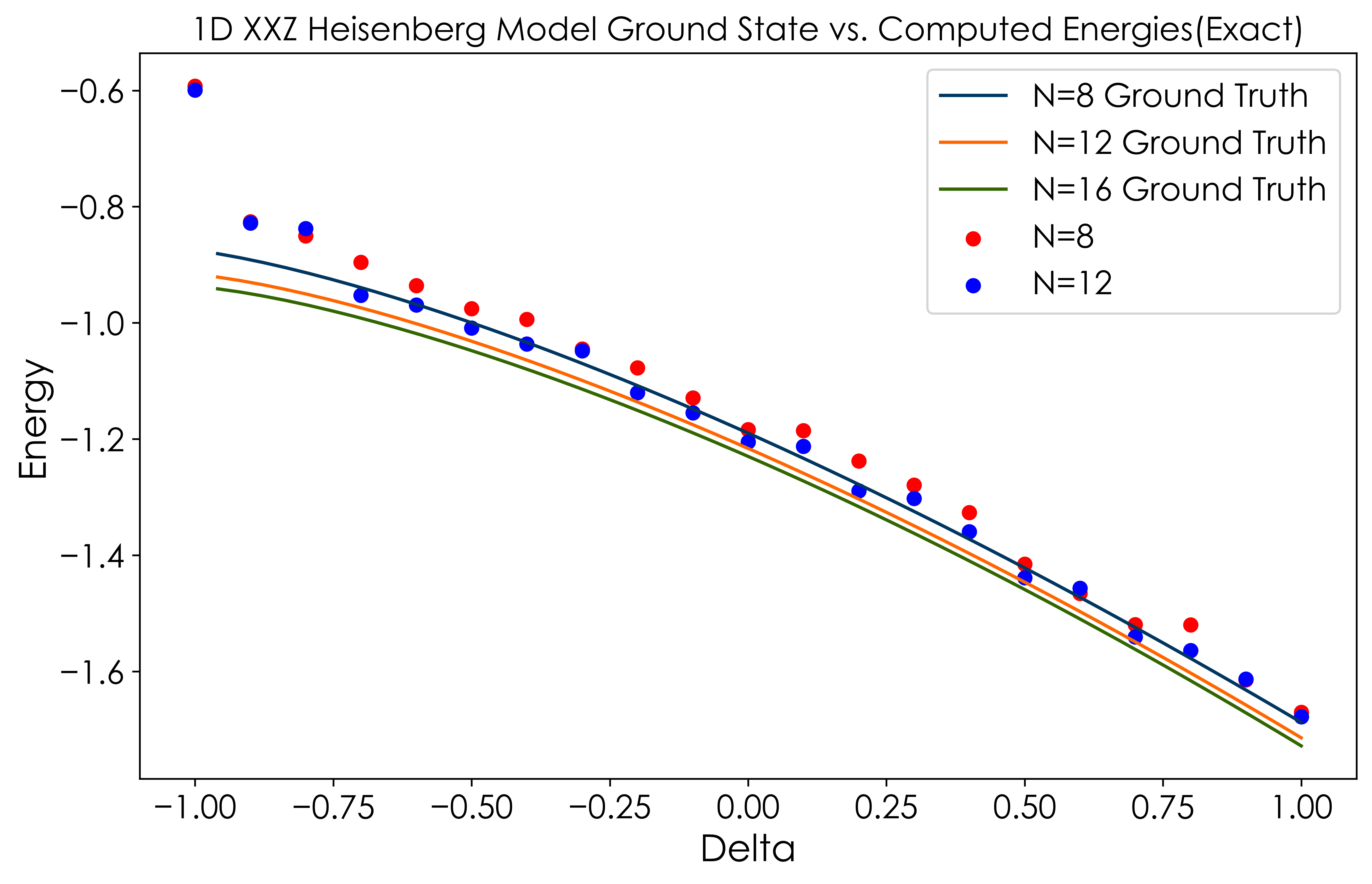}
    \caption{Ground states of the 1D Anisotropic XXZ Heisenberg model with  in the range of [-1,1] sampled at an interval length of 0.1, classically computed expectation value}
    \label{fig:apd2}
\end{figure}

\subsection{Prepared State Verification}

\subsubsection{Correlation Function}
The XXZ model is in the critical phase when the anisotropy parameter satisfies $ -1 \leq \Delta \leq +1$. This is characterized by long-range correlations (polynomially decaying correlations). For $ \Delta \geq 1 $, the model is in a gapped phase characterized by exponentially decaying correlations. These trends can be seen from the correlation function plots below.

\begin{figure}[h]
    \centering
    \includegraphics[width=0.75\textwidth]{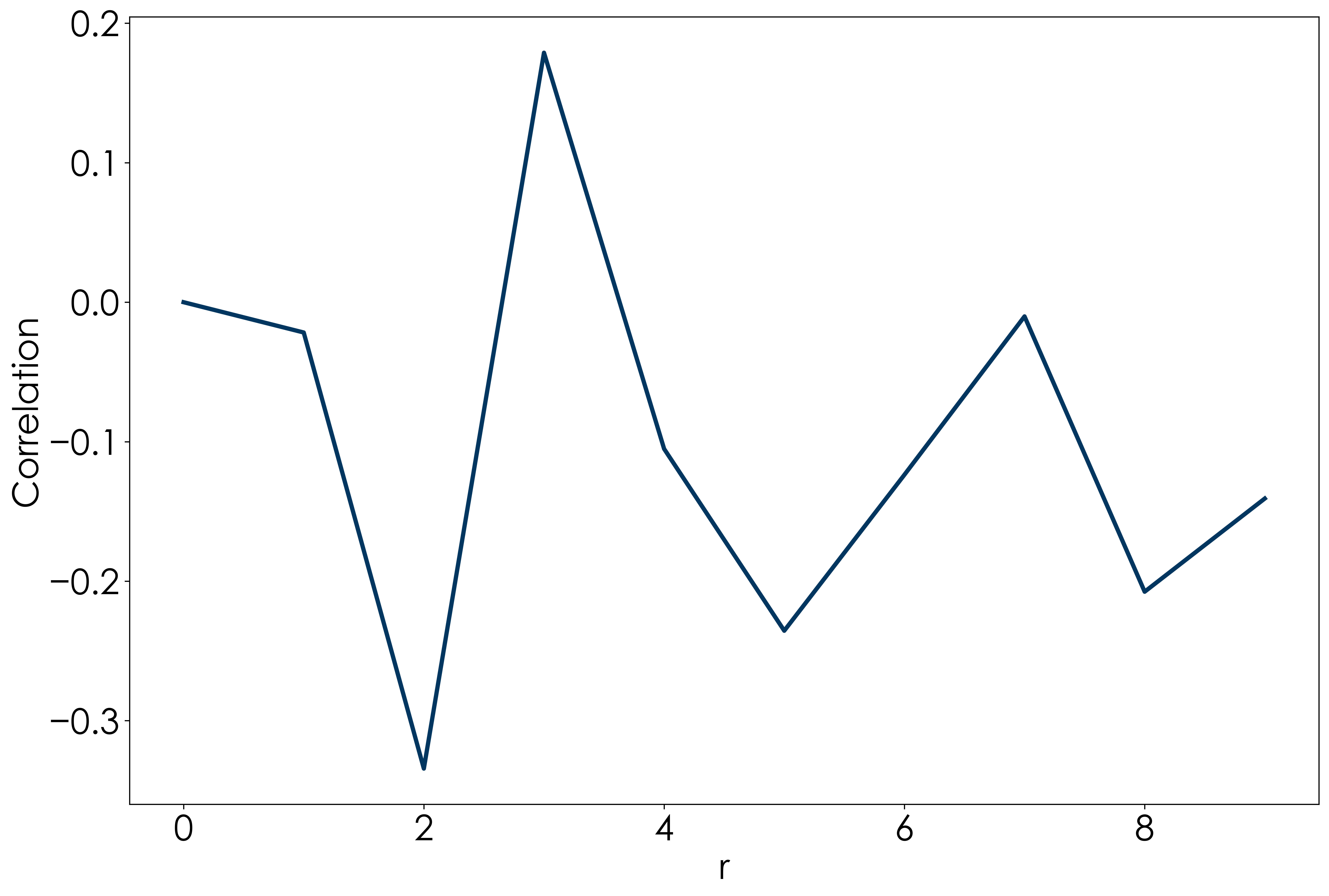}
    \caption{the correlation of the prepared state for $N=10$ and $\Delta=0.3$ with respect to r(distance between qubits)}
    \label{fig:apd3}
\end{figure}

\begin{figure}[h]
    \centering
    \includegraphics[width=0.75\textwidth]{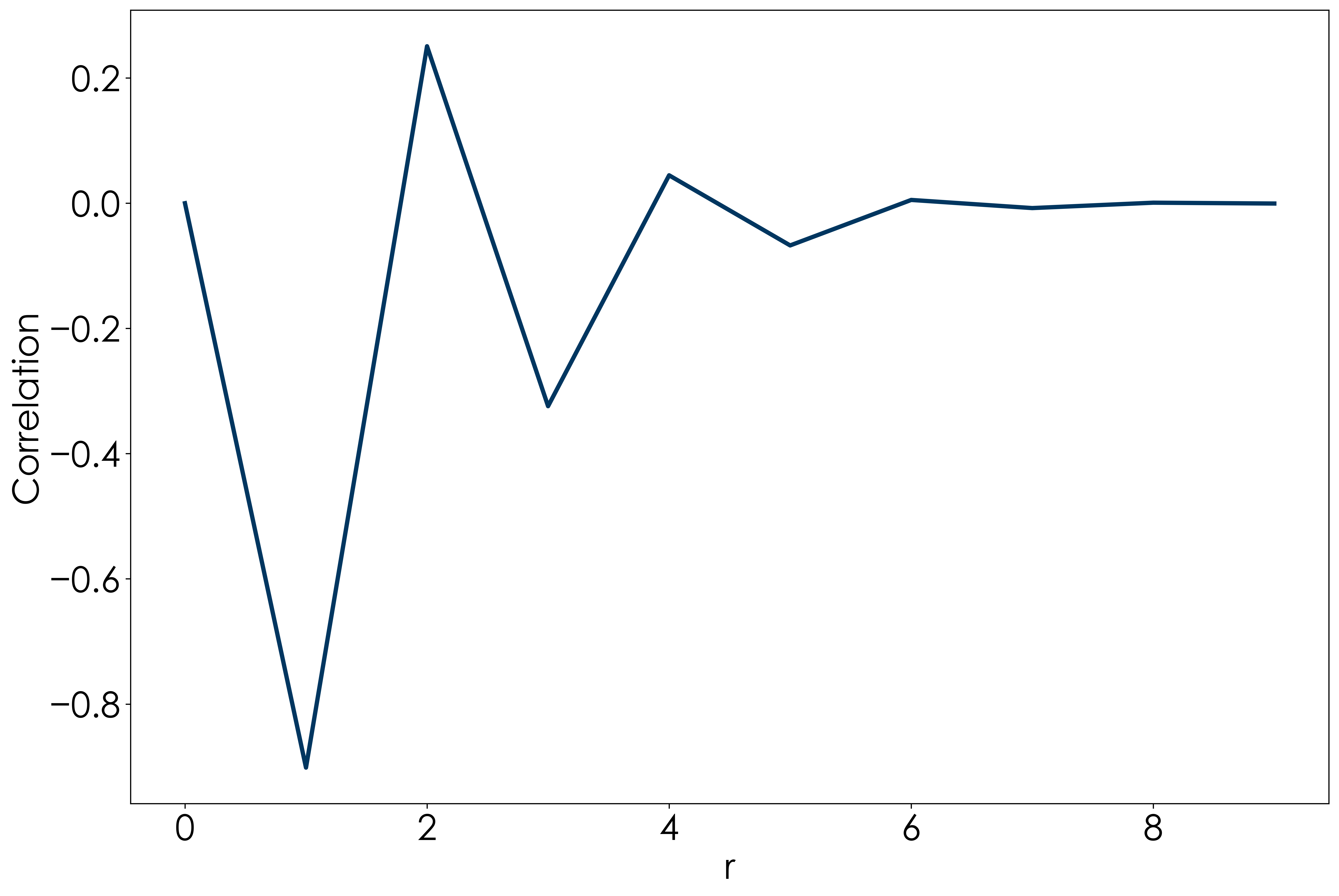}
    \caption{the correlation of the prepared state for $N=10$ and $\Delta=1.0$ with respect to r}
    \label{fig:apd4}
\end{figure}

\begin{figure}[h]
    \centering
    \includegraphics[width=0.75\textwidth]{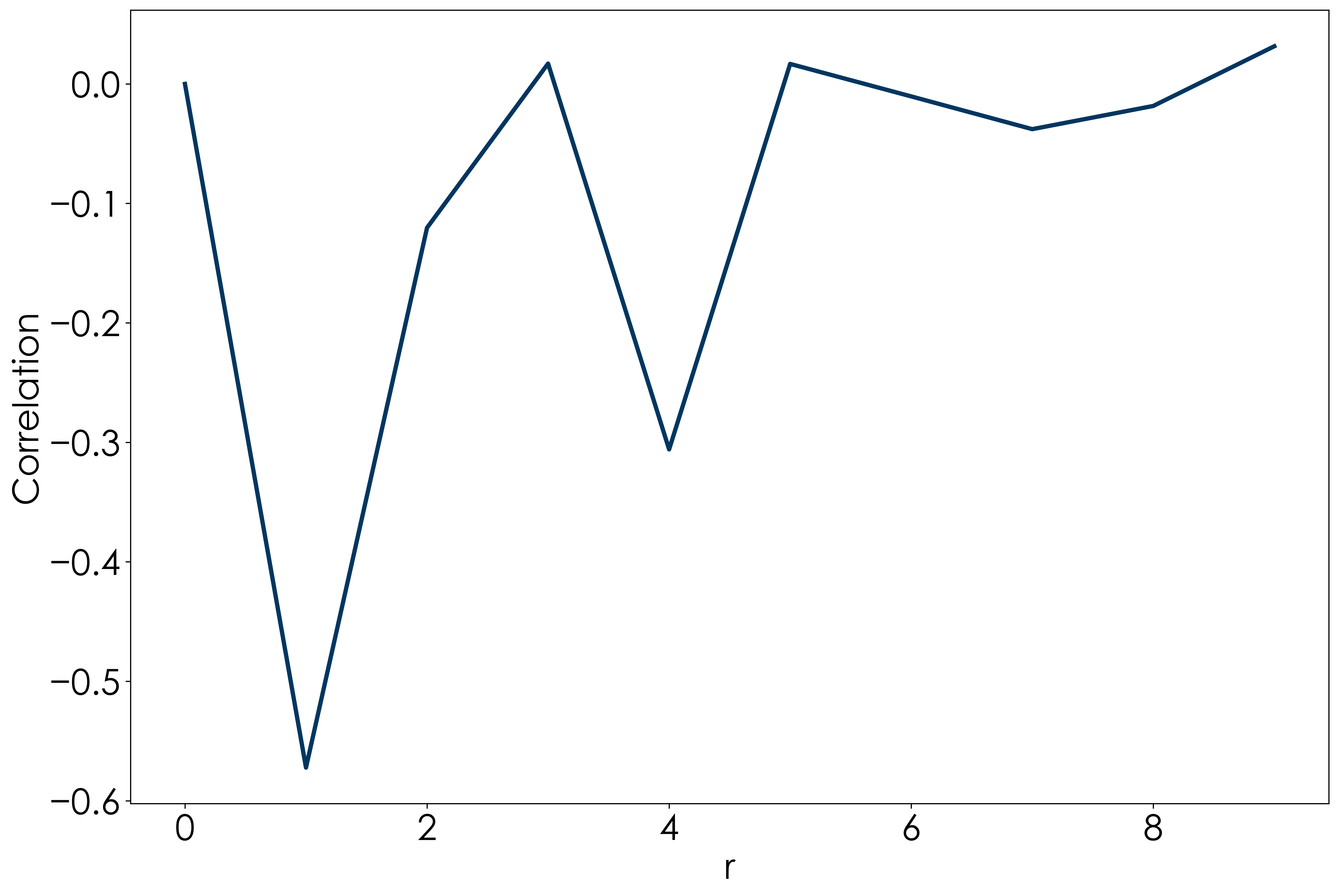}
    \caption{the correlation of the prepared state for $N=10$ and $\Delta=2.5$ with respect to r}
    \label{fig:apd5}
\end{figure}

\subsubsection{Subsystem Entropy}
For $\Delta=1$ , the subsystem entropy displays an alternating profile characteristic of the dimer-like nature of the ground state of the Heisenberg model. This vanishes as one goes away from the isotropic limit.

\begin{figure}[h]
    \centering
    \includegraphics[width=0.75\textwidth]{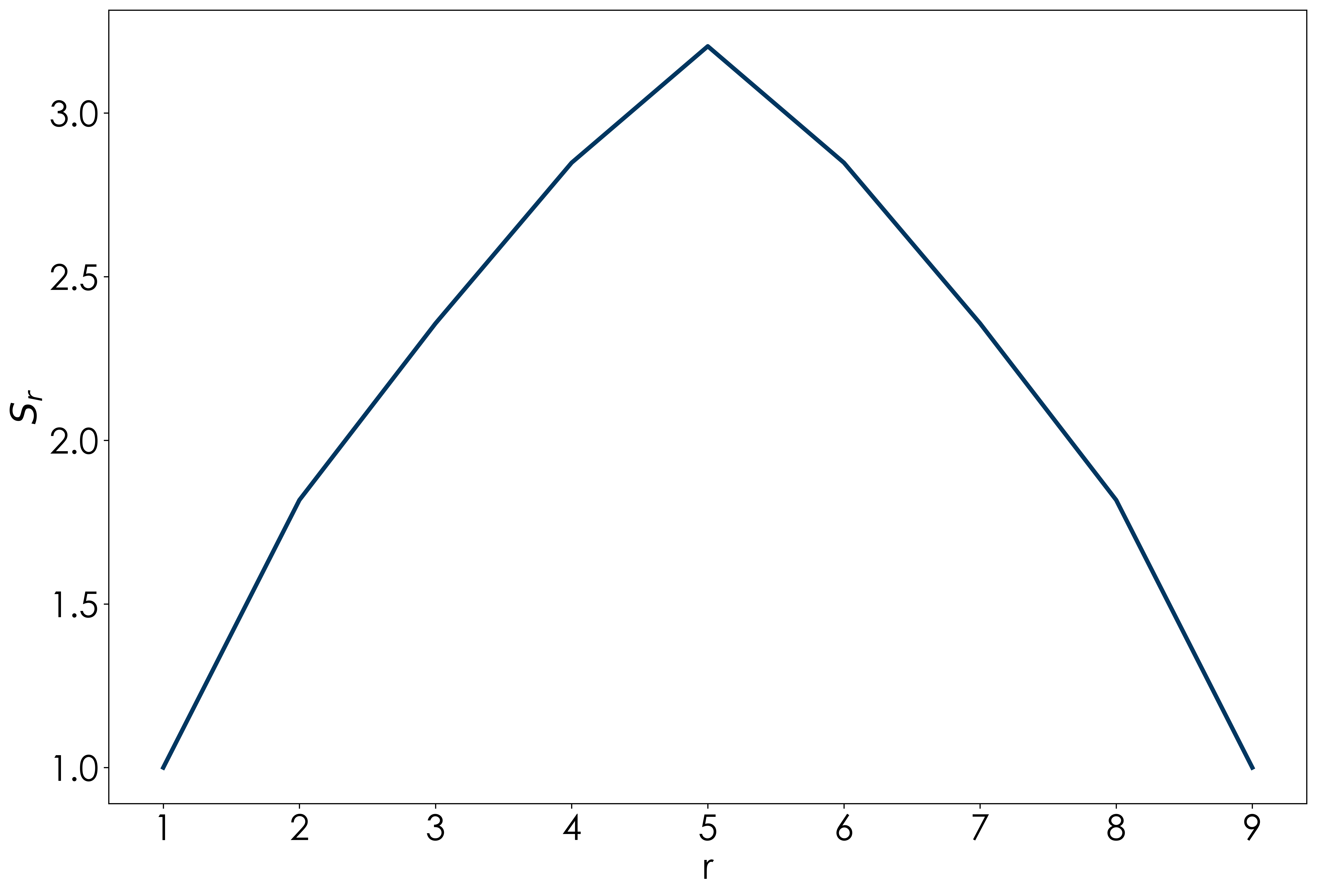}
    \caption{the subsystem entropy of the prepared states for $N=10$ and $\Delta=0.3$ }
    \label{fig:apd6}
\end{figure}

\begin{figure}[h]
    \centering
    \includegraphics[width=0.75\textwidth]{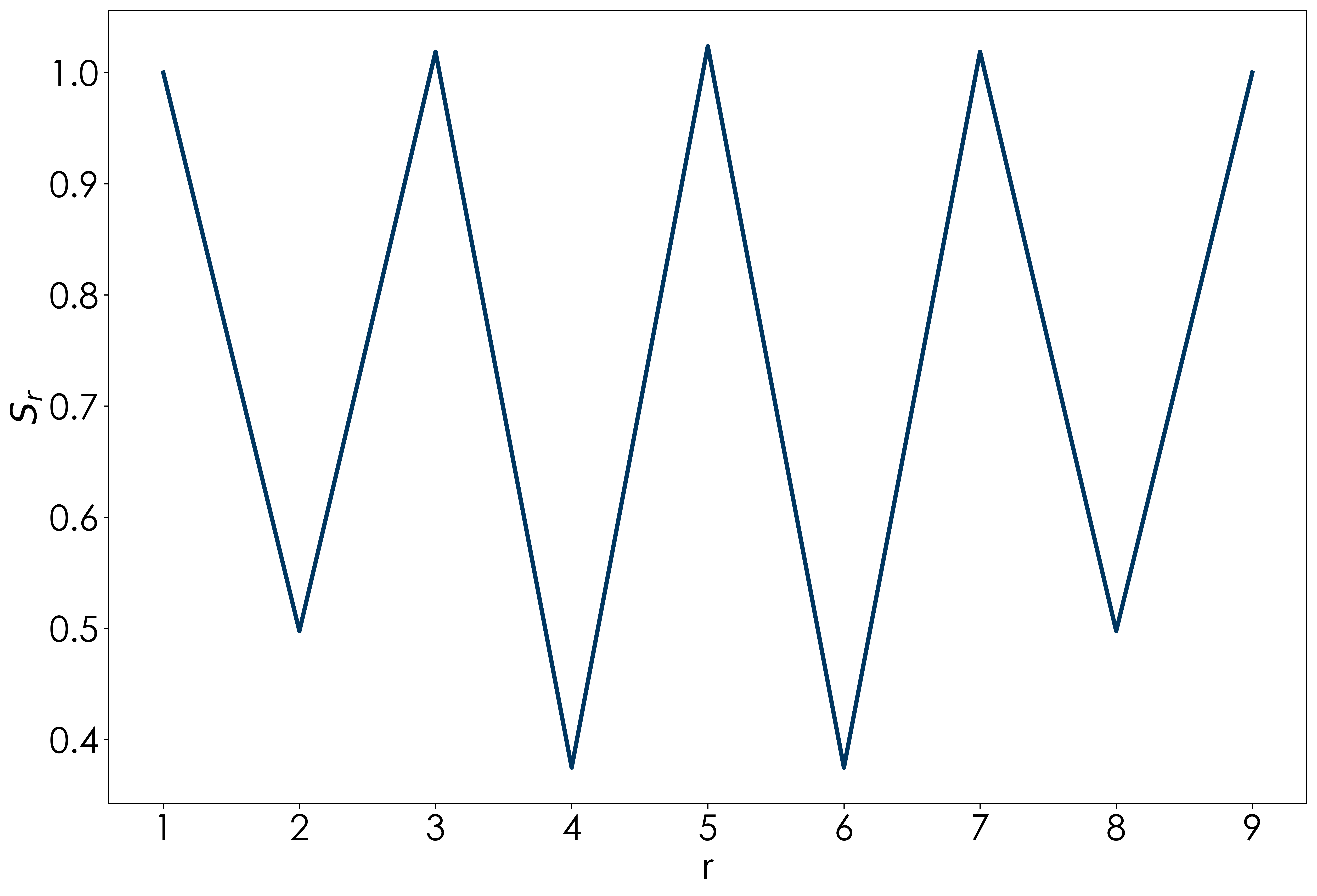}
    \caption{the subsystem entropy of the prepared states for $N=10$ and $\Delta=1.0$ }
    \label{fig:apd7}
\end{figure}

\begin{figure}[h]
    \centering
    \includegraphics[width=0.75\textwidth]{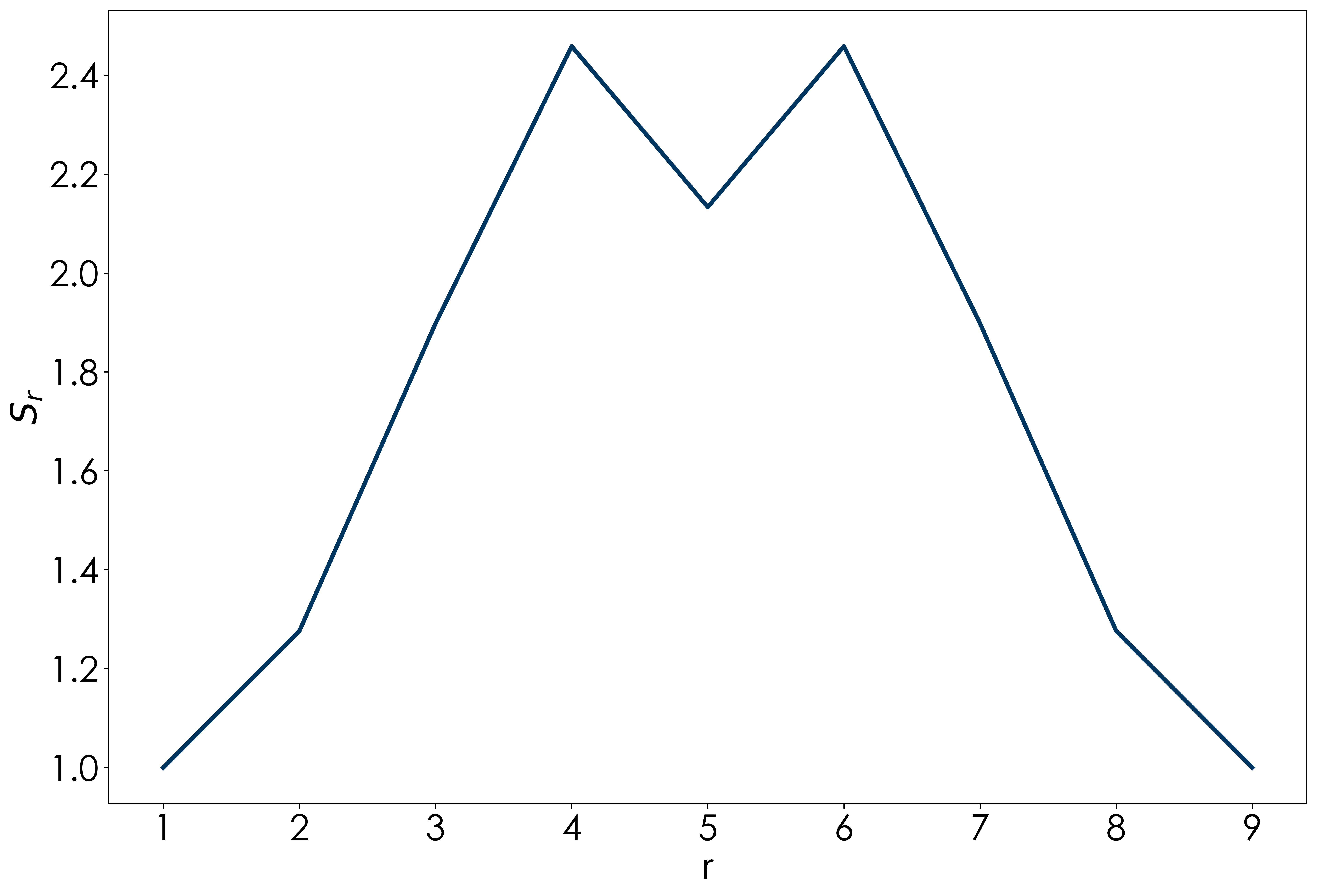}
    \caption{the subsystem entropy of the prepared states for $N=10$ and $\Delta=2.5$ }
    \label{fig:apd8}
\end{figure}

\end{document}